\documentclass[12pt]{elsarticle}

\makeatletter
\def\ps@pprintTitle{%
 \let\@oddhead\@empty
 \let\@evenhead\@empty
 \def\@oddfoot{\centerline{\thepage}}%
 \def\@evenfoot{\thepage}
 \let\@evenfoot\@oddfoot}
\makeatother

\usepackage[T1]{fontenc}
\usepackage{lmodern}
\usepackage{dirtytalk}
\usepackage{microtype}
\usepackage{comment}
\usepackage[margin=1in]{geometry}
\usepackage[]{algorithm2e}
\usepackage{amsmath,amssymb,amsthm,bm}
\usepackage{mathtools}
\usepackage{booktabs}
\usepackage{graphicx}
\usepackage{pdfpages}
 \usepackage{algpseudocode} 
\usepackage{subcaption}
\biboptions{sort&compress}
\usepackage{epstopdf}
\usepackage{float}
\usepackage{miller}
\graphicspath{ {./Photos/} }

\usepackage[utf8x]{inputenc}
\usepackage{xcolor}

\makeatletter
\def\fnmark[#1]{\def\@fnmark{$\dagger$}}
\def\fntext[#1]#2{\g@addto@macro\@fnotes{%
  \refstepcounter{fnote}\elsLabel{#1}%
  \def\thefootnote{$\dagger$}%
  \footnotetext{#2}}}
\makeatother

\AppendGraphicsExtensions{.tif}

\begin{document}
\begin{frontmatter}

\title{Dislocation mediated twin boundary faceting in $\beta$-Sn}

\author[XCP5,IA]{John Israel\fnref{cofirst}}
\author[XCP5]{Ian Chesser\corref{cor1}\fnref{cofirst}}
\author[MST]{Khanh Dang}
\author[XCP5]{Vinamra Agrawal}

\cortext[cor1]{Corresponding author. \textit{Email address:} ichesser@lanl.gov}
\fntext[cofirst]{These authors contributed equally to this work.}

\address[XCP5]{XCP-5, Los Alamos National Laboratory, Los Alamos, NM USA}
\address[IA]{Department of Aerospace Engineering, Iowa State University, Ames, IA USA}
\address[MST]{MST-8, Los Alamos National Laboratory, Los Alamos, NM USA}

\begin{abstract}
    Grain boundaries (GBs), including twin boundaries (TBs), play a critical role in plastic deformation by interacting with dislocations. The deformation and strengthening mechanisms in $\beta$-Sn are of significant interest because Sn-based alloys are widely used as solder materials. Recent experiments have shown deformation-induced faceting in $\beta$-Sn arising from twin-dislocation interactions. Using molecular dynamics (MD) simulations informed by a recently developed machine-learned interatomic potential (MLIP), we demonstrate a \hkl<c> dislocation-absorption mediated faceting reaction along \hkl(301) TBs in $\beta$-Sn in which facet length grows with the number of absorbed dislocations. We show that the predicted facet length is consistent with isotropic elasticity theory only when the disclination character of the asymmetric twin facet is taken into account. We discuss analogies to faceting reactions in other low symmetry metals and hypothesize that the ubiquity of \hkl(301) twins in $\beta$-Sn stems from their ability to grow rapidly through a forest of dislocation obstacles. 

\end{abstract}


\end{frontmatter}



%
At room temperature, Sn exists in a low symmetry body centered tetragonal (BCT) phase called $\beta$-Sn with a $c/a$ ratio of 0.55. Alloyed $\beta$-Sn is used extensively in lead-free solder joints in the electronics industry where its deformation behavior and failure directly impacts the reliability of electronic components \cite{osorio2013microstructure,lehman2010cyclic}. A typical deformation path in $\beta$-Sn, such as that experienced during bending-induced 'tin-cry' \cite{chalmers1932cry}, involves a combination of dislocation-mediated and twin-mediated plasticity \cite{minor2014}. Twinning is an important deformation mechanism in $\beta$-Sn due to the prevalence of low energy twin boundaries (TBs) in processed Sn microstructures and due to the lack of easily activated dislocation slip systems in $\beta$-Sn, especially at low temperatures and high strain rates \cite{tu1970direct, yang2007deformation, minor2014, pokharel2023, nguyen2024calibration, PTW2025}. 

A surprising observation from recent $\beta$-Sn bicrystal compression experiments \cite{kaira} is that, contrary to conventional wisdom, \hkl(301) TBs were found to soften the bicrystal in comparison to single crystal specimens of equivalent orientation. The \hkl(301) twin system is the most commonly observed twin misorientation across deformation and solidification experiments in $\beta$-Sn \cite{rowenhorst2005measurements, telang2005orientation, kaira2016microscale,  reeve2018beta, gupta2022electrodeposition}. Rather than acting as an obstacle to slip, \hkl(301) TBs were found to accommodate dislocations within the TB structure through a combination of facet formation, TB migration and partial transmission of dislocations \cite{kaira}. Relatively large asymmetric twin facets of up to 50 nm were observed along an initially flat TB after deformation in $\beta$-Sn, reminiscent of asymmetric basal--prismatic (BP) and prismatic--basal (PB) twin facets observed in hexagonal close-packed (HCP) metals such as Mg, Ti or Zr after deformation \cite{gong2018basal,hooshmand2017atomistic,yue2023twin, barrett2014impact}. Atomistic simulations have shown that, in HCP metals, dislocation-absorbing asymmetric facets allow rapid twin growth through a forest of dislocation obstacles, helping to explain the ubiquity of \hkl(10-12) twins in Mg \cite{barrett2014faceting,barrett2014roles,el2015}. Dislocation--twin interactions in $\beta$-Sn are still poorly understood relative to metals like Mg, although recent atomistic simulations have demonstrated the existence of asymmetric A-plane--Prismatic (AP) facets during \hkl(301) twin growth in the absence of lattice dislocations \cite{Ian}.

In this work, we investigate the interaction between dislocations and the \hkl(301) coherent twin boundary (CTB) in $\beta$-Sn using molecular dynamics (MD) simulations informed by a machine-learned interatomic potential (MLIP). We consider the \hkl{100}\hkl<001> slip system, whose dislocations are denoted \hkl<c> dislocations for short, and which is commonly observed in deformation experiments \cite{yang2007deformation,kaira,Bhatia_2017}. Over a range of applied stresses and temperatures, we observe that \hkl<c> dislocations are absorbed into mobile TBs to form asymmetric AP facets that lengthen with the number of absorbed dislocations. We demonstrate that this reaction is consistent with predictions from isotropic elasticity only when the disclination character of the AP facet is taken into account. Finally, we discuss this reaction relative to the experimental deformation-induced faceting results of \cite{kaira} and compare and contrast it to twin--dislocation reactions observed in HCP metals. 

Simulations were conducted using the Large-scale Atomic/Molecular Massively Parallel Simulator (LAMMPS) \cite{LAMMPS} and the moment tensor potential (MTP) MLIP for $\beta$-Sn developed in \cite{Ian}. This potential accurately captures stacking-fault energies for various slip systems compared with density functional theory (DFT), with extensive validation of twinning behavior against experiments in \cite{Ian}. We note that, despite its strengths, this potential significantly underestimates the melting point of $\beta$-Sn (approximately 300 K compared with 505 K experimentally) and inherits an overly soft $C_{44}$ elastic constant from DFT (8 GPa compared with 21.9 GPa). The Open Visualization Tool (OVITO) \cite{OVITO} was used to visualize atomic structures and dynamics with bulk BCT coordinated atoms identified by the polyhedral template matching (PTM) modifier \cite{PTM} (root-mean-square deviation cutoff of 0.25) via approximately simple-cubic coordination. The near simple-cubic coordination of bulk atoms in $\beta$-Sn is a consequence of its multi-lattice structure of interpenetrating BCT lattices separated by a translation of \hkl[a/2, 0, c/4]. 

\begin{figure}[htb]
    \centering
    \includegraphics[width=0.7\linewidth]{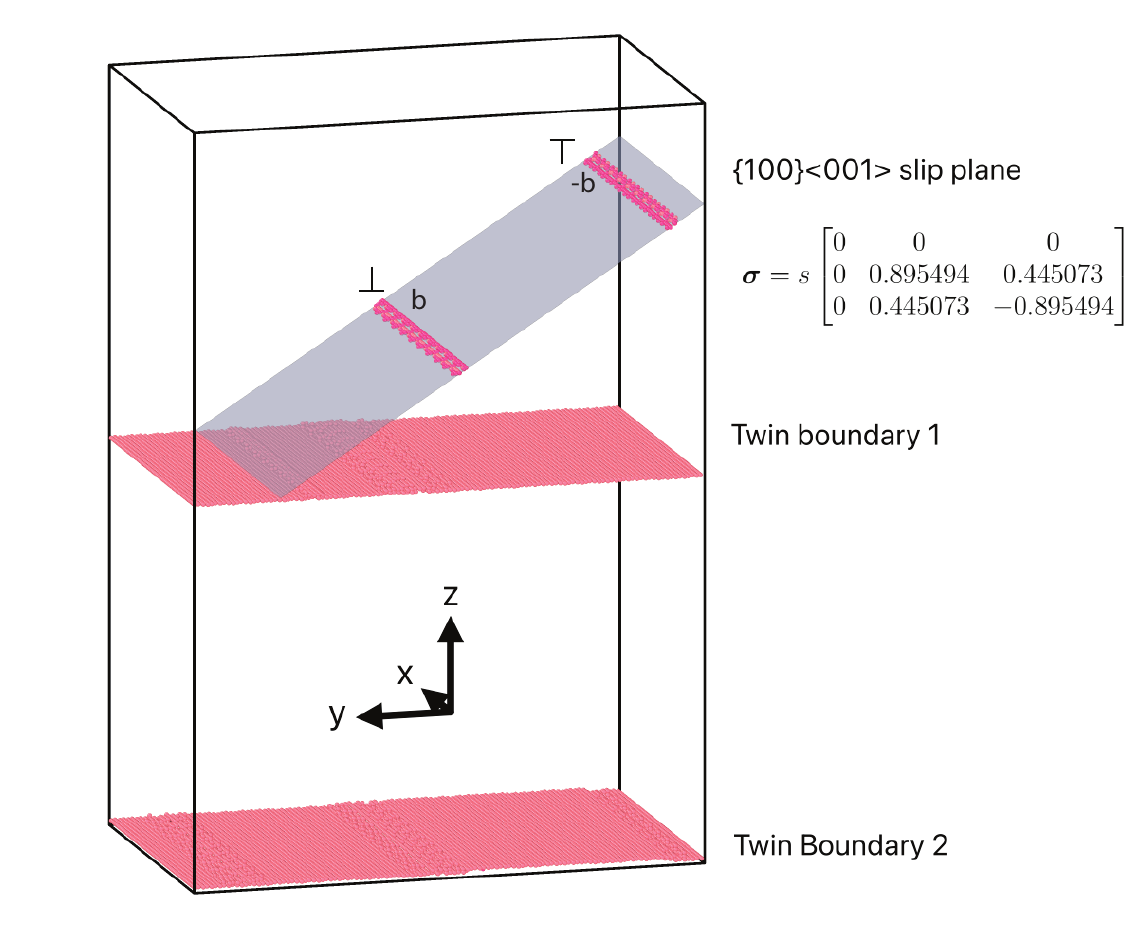}
    \caption{Simulation setup of a periodic bicrystal containing two \hkl(301) TBs and a \hkl<c>-type edge dislocation dipole. A pure shear stress $s$ resolved along the slip plane is given in the simulation frame.}
    \label{fig:schematic}
\end{figure}

Relaxed atomistic \hkl(301) TB structures were generated via a grand-canonical optimization routine implemented in the open-source GRIP code \cite{GRIP}, which samples a more diverse set of grain boundary (GB) densities than the traditionally used $\gamma$-search method \cite{olmsted2009survey}. These structures were described previously in Ref.~\cite{Ian} and were resized in this work to produce bicrystals with dimensions $4.2\times68\times100$ nm$^3$ (990K atoms), where $X$ is parallel to the \hkl[010] tilt axis, the $XY$ plane contains the TB, and $Z$ is normal to the TB (Fig.~\ref{fig:schematic}). A \hkl<c> edge-dislocation dipole was inserted in the upper grain, which we denote the twin grain, following the methodology described in Ref.~\cite{dang2017lammpsedit}. To study interactions of a TB with multiple dislocations, nested dipoles were inserted on the same slip plane with sufficient spacing to avoid rapid annihilation during minimization or equilibration. For the grain thickness of 50 nm, four dipoles could be inserted without self-annihilation. 

After a short equilibration of the dislocation--TB system in the isothermal--isobaric (NPT) ensemble at zero pressure, a stress tensor was applied that resolves a pure shear stress on the dislocation slip plane. In the simulation frame, as shown in Fig.~\ref{fig:schematic}, this stress state exerts a positive shear stress in the $YZ$ plane, a tensile stress along $Y$, and a compressive stress along $Z$. Two driving forces exist for shear-coupled TB migration: a Peach--Koehler force due to the $YZ$ shear stress resolved on the boundary and an elastic-energy-density jump across the TB driven by the normal stresses. The former dominates the latter, with a driving force for shear-coupled migration $|\sigma_{yz}\beta|=0.0525s$, where $s$ is the shear stress resolved on the \hkl{100}\hkl<001> slip system in the twin grain and $\beta$ is the shear-coupling factor of the \hkl(301) twin (approximately $-0.1$ \cite{Ian}). The negative coupling factor means that the TB migrates downward (-Z) in response to a positive shear (+YZ). Simulations were run over a grid of temperature and stress values for temperatures of 10, 50, 100, 150 and 200 K and for $s$ values of 50 to 500 MPa in increments of 50 MPa for a total simulation time of 0.1 ns. Statistics for TB-dislocation interactions were averaged over the two TBs interacting with one dislocation each per inserted dipole. Simulations with multiple dislocation dipoles were conducted only at 100 K and $s = 250$ MPa. 

\begin{figure}[t]
    \centering
    \includegraphics[width=\textwidth]{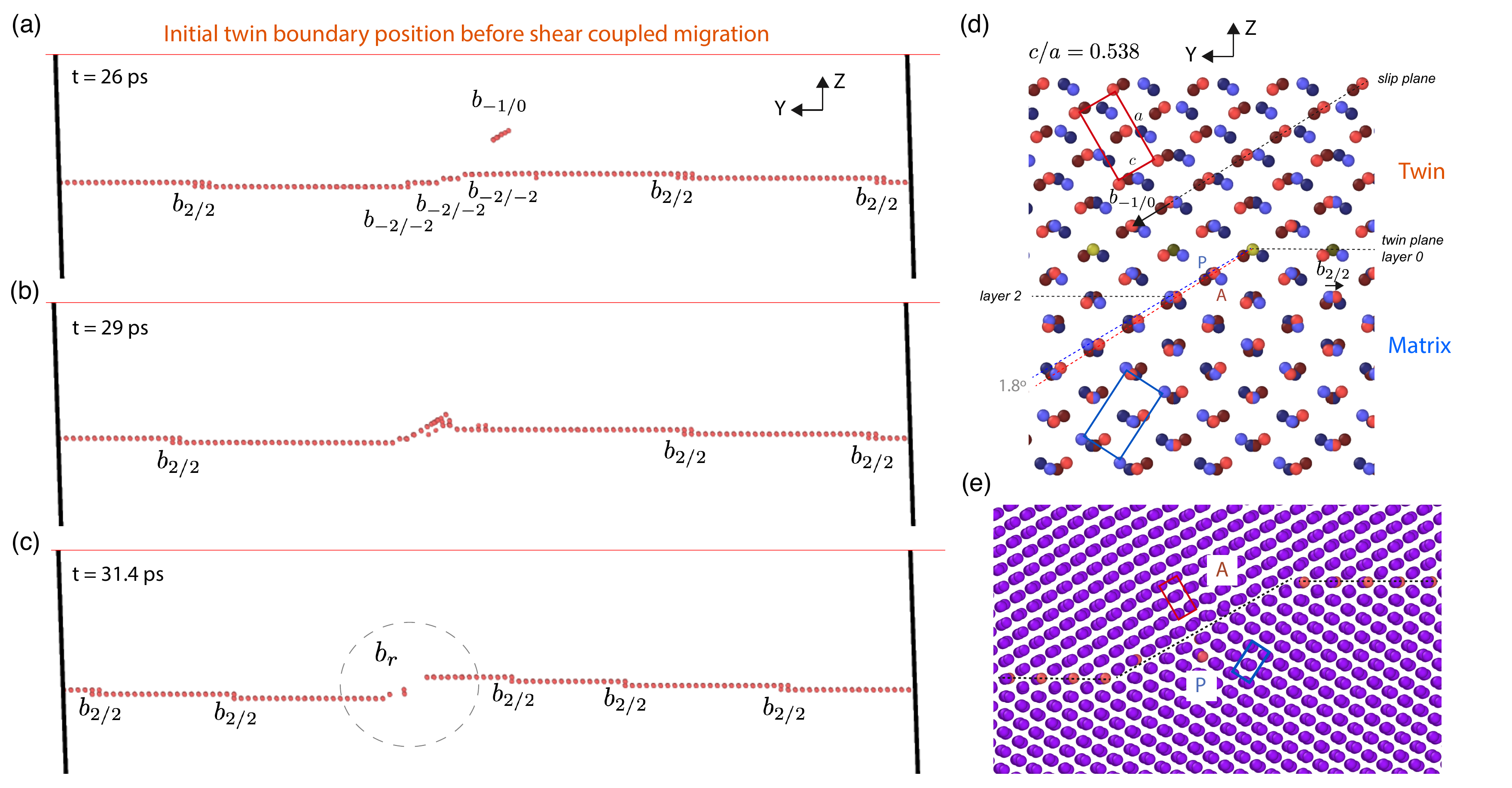}
    \caption{(a)-(c) AP facet forming dislocation absorption reaction along with (d) relevant bi-crystallography and (e) the atomic structure of the AP facet. In (a)-(c), burgers vectors of relevant defects are labeled. In (d)-(e), red and blue rectangles show the $\beta$-Sn unit cell in the twin and matrix grains.}
    \label{fig3:trajectory}
\end{figure}

In our simulations, \hkl{100}\hkl<001> dislocation glide begins between 100 and 150 MPa, after the onset of shear coupled TB migration. Once the dislocation becomes mobile, it chases the migrating TB to become absorbed in the boundary at stresses up to 450 MPa (Fig.~\ref{fig3:trajectory}a-c). This facet-forming reaction is the primary reaction of interest in this work. At stresses between 450 and 500 MPa, nucleation and growth of a secondary twin inclusion are observed at the pre-existing TB, with a misorientation consistent with \hkl{301} twinning \cite{Ian}. This heterogeneous twin-nucleation mechanism provides a possible recrystallization pathway in Sn and is discussed further in the Supplementary Information. Related dislocation-assisted twin-nucleation and twin-growth mechanisms have been analyzed in HCP Mg and Ti \cite{ghazisaeidi2013dissociation,hooshmand2017atomistic,Wang01092013}. 

The facet forming absorption reaction is shown in detail in Fig.~\ref{fig3:trajectory}a-c along with crystallographic defects relevant to the reaction in Fig.~\ref{fig3:trajectory}d-e. Before the dislocation impinges on the TB, three unit twinning disconnection (TD) dipoles decorate the mobile TB (Fig.~\ref{fig3:trajectory}a). The negative signed disconnections coalesce to form a facet stabilized by the incoming dislocation (Fig.~\ref{fig3:trajectory}b-c). The facet is an asymmetric AP facet \cite{Ian}, bounded by \hkl{100} A-type planes in the twin grain and \hkl{101} P-type planes in the matrix grain (Fig.~\ref{fig3:trajectory}d-e). The A and P planes are mis-aligned in the dichromatic pattern as shown by the rotation of $1.8^\circ$ for the MTP lattice parameters in Fig.~\ref{fig3:trajectory}d, implying that the Burgers vector content of the AP facet can be partitioned into a rotational component associated with a disclination dipole $\mathbf b_{r, AP}$ and a translational component $\mathbf b_{t, AP}$ associated with the coalesced disconnections.

The absorption reaction can be written as

\begin{equation}
\mathbf b_{-1/0}
=
\mathbf b_{t, AP}
+
\mathbf b_{r,AP}
+
n\mathbf b_{2/2}.
\end{equation}

 $\mathbf b_{-1/0}$ denotes the incoming dislocation in the twin grain and $n \mathbf b_{2/2}$ the Burgers-vector content of the TDs left behind at the TB. We use the $b_{p/q}$ notation of Serra and Bacon \cite{serra1996new}, where $p$ and $q$ count the number of \hkl(301) twin planes terminated by the Burgers vector in the twin and matrix grains and $-b_{p/q}=b_{-p/-q}$. Note that $b_{-1/0}$ and $b_{2/2}$ have opposite signs along $Y$ in the dichromatic pattern (Fig.~\ref{fig3:trajectory}d). Conservation of the $p/q$ indices implies $\mathbf b_{t,AP}+\mathbf b_{r,AP}=\mathbf b_{-(2n+1)/-2n}$. For $n=5$ TDs participating in the reaction, as observed in Fig.~\ref{fig3:trajectory}c, the $\mathbf b_{-(2n+1)/-2n}$ disconnection is predicted to span 11 \hkl(301) planes in the twin grain and 10 \hkl(301) planes in the matrix grain, consistent with the AP-facet structure observed in Fig.~\ref{fig3:trajectory}e.   

An estimate for the AP-facet length per absorbed dislocation can be obtained from isotropic elasticity theory by treating $n$ as a continuous variable and minimizing the elastic energy of the translational residual and the $n$ isolated TDs produced by the reaction with respect to n. Assuming a common prefactor $C$ for elastic energies, the minimization becomes 
\begin{equation}
E(n)
=
C\left[
\left|
\mathbf b_{-1/0} - \mathbf b_{r,AP} - n\mathbf b_{2/2}
\right|^2
+
n\left|
\mathbf b_{2/2}
\right|^2
\right],
\qquad
\frac{dE}{dn} = 0
\label{eq:AP_reaction_energy}
\end{equation}

The value that minimizes Eq.~\eqref{eq:AP_reaction_energy} is denoted $n^*$,
with its derivation given in the Supplementary Information.  Once $n^*$ is
known, the corresponding facet length follows directly from the step height of
the residual $\mathbf b_{-(2n^*+1)/-2n^*}$.  The step height of a general
$b_{p/q}$ defect is
\begin{equation}
h_{p/q}
=
\frac{p+q}{2}d_{301},
\qquad
d_{301}
=
\frac{ac}{\sqrt{a^2+9c^2}}.
\end{equation}
The magnitude of the residual step height is therefore
\begin{equation}
H_r(n^*)
=
2d_{301}\left(n^*+\frac{1}{4}\right).
\label{eq:residual_height_nstar}
\end{equation}
Projecting this height onto the exact A- and P-plane directions gives,
respectively,
\begin{align}
L_A(n^*)
&=
2c\left(n^*+\frac{1}{4}\right),\\
L_P(n^*)
&=
\sqrt{a^2+c^2}
\left(n^*+\frac{1}{4}\right).
\label{eq:AP_lengths_nstar}
\end{align}
Because the A and P planes are slightly misaligned, these expressions provide
separate twin-side and matrix-side estimates of the AP-facet length.

Equation~\eqref{eq:AP_reaction_energy} must be solved self-consistently because
the Burgers vector content of the disclination dipole $\mathbf b_{r,\mathrm{AP}}$ depends on
the facet length, whereas the facet length itself depends on $n^*$ through the
residual step height.  Approximating the AP-facet length by the mean of the A
and P-side lengths gives for the MTP lattice parameters $a=5.97\,\text{\AA}$ and
$c=3.22\,\text{\AA}$ the self-consistent solution $n^*=4.98$
which predicts the nearest-integer reaction $n=5$ in good agreement with our simulations. The experimental lattice parameters $a=5.83\,\text{\AA}$ and
$c=3.18\,\text{\AA}$ give $n^*=6.14$ corresponding to the
nearest-integer prediction $n=6$. As shown in the Supplementary Information, the inclusion of the disclination content in the minimization problem lowers the predicted $n$ and is important to obtaining the good agreement observed between simulation and theory. 

\begin{figure}[ht!]
    \centering
    \begin{subfigure}[h]{0.23\textwidth}
        \centering
        \includegraphics[width=\textwidth]{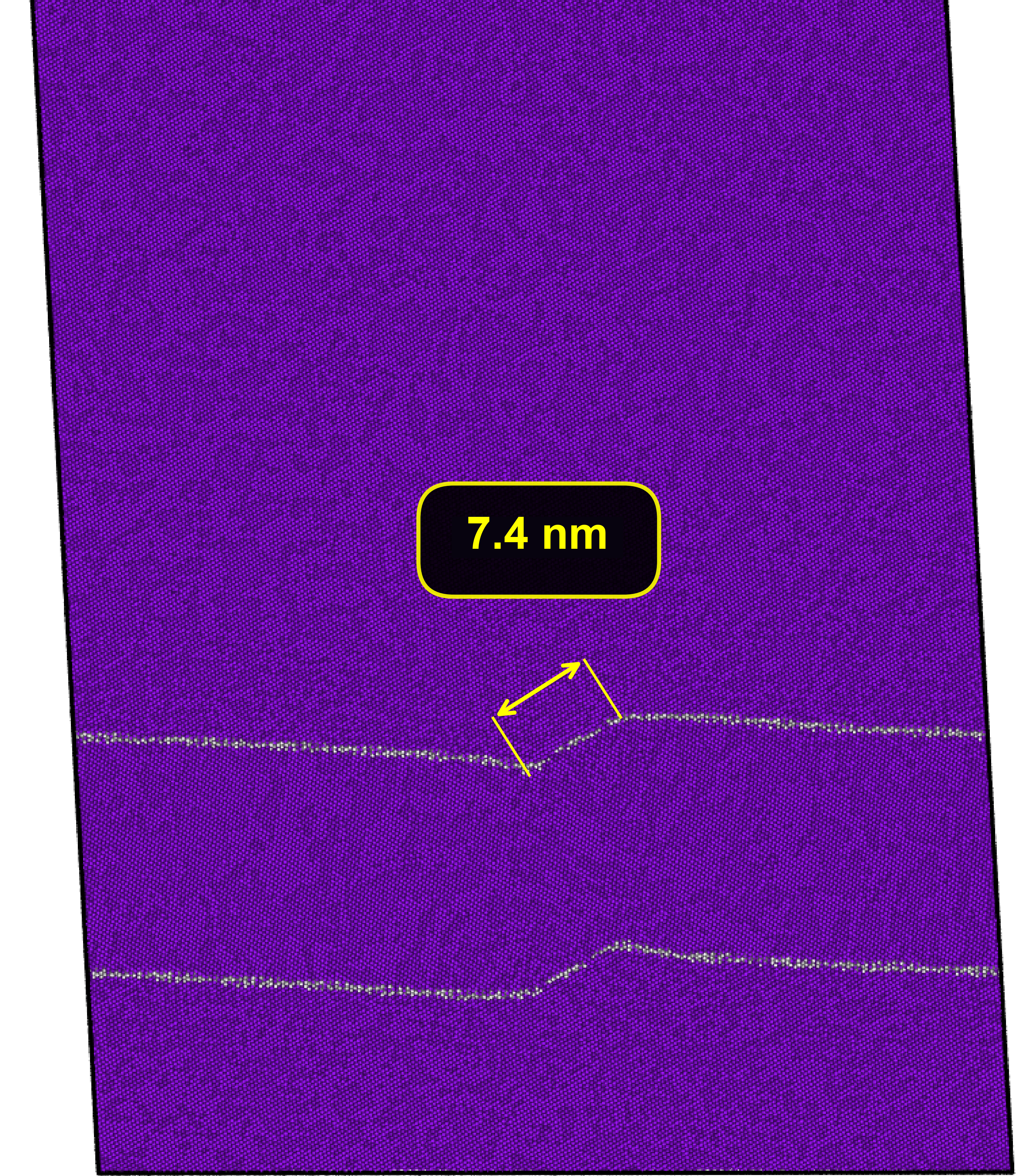}
        \caption{Facet with 2 Dislocations Absorbed}
        \label{fig:sub2}
    \end{subfigure}
    \hfill
    \begin{subfigure}[h]{0.23\textwidth}
        \centering
        \includegraphics[width=\textwidth]{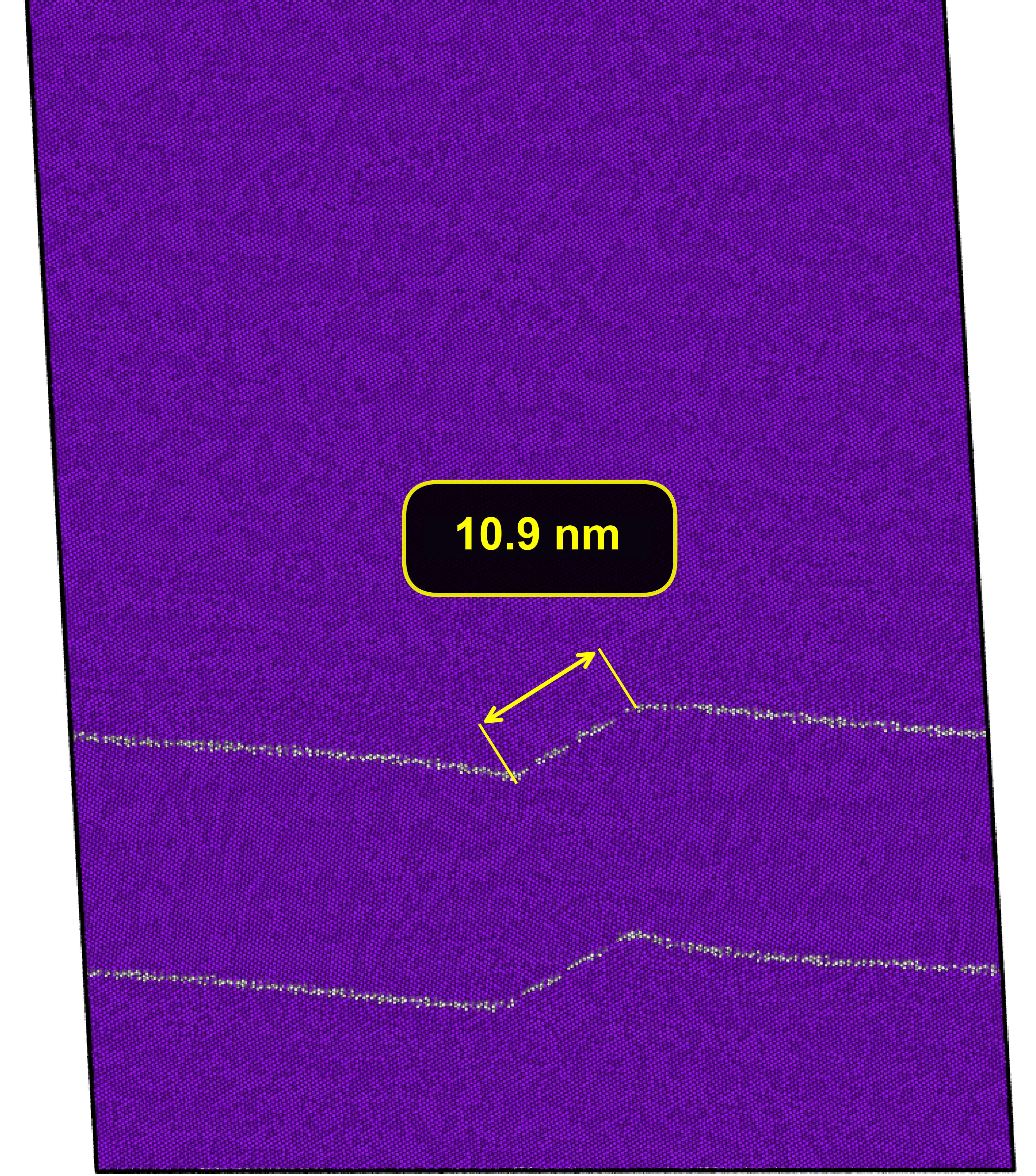}
        \caption{Facet with 3 Dislocations Absorbed}
        \label{fig:sub3}
    \end{subfigure}
    \hfill
    \begin{subfigure}[h]{0.23\textwidth}
        \centering
        \includegraphics[width=\textwidth]{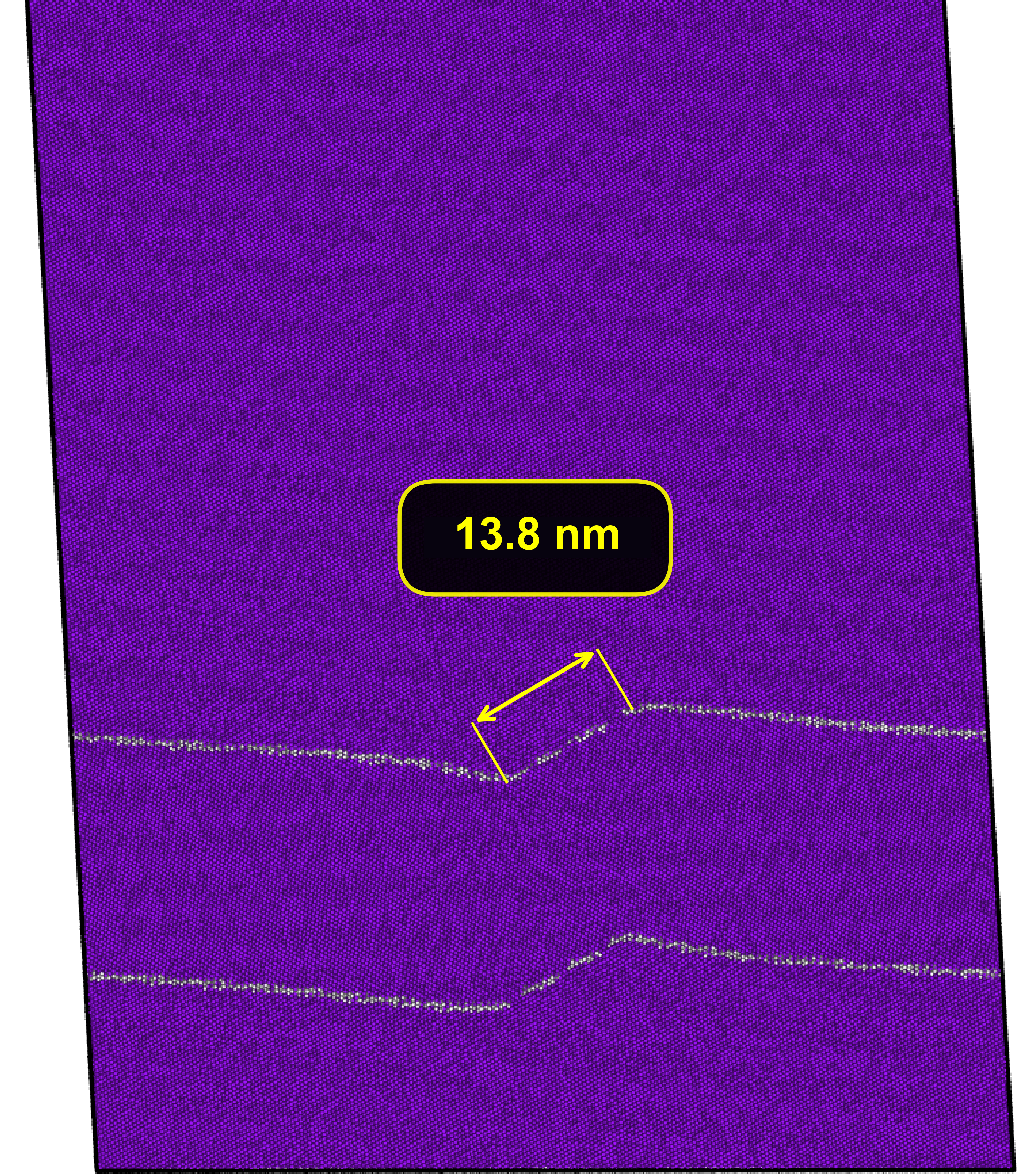}
        \caption{Facet with 4 Dislocations Absorbed}
        \label{fig:sub4}
    \end{subfigure}
    \hfill
    \begin{subfigure}[h]{0.23\textwidth}
        \centering
        \includegraphics[width=\textwidth]{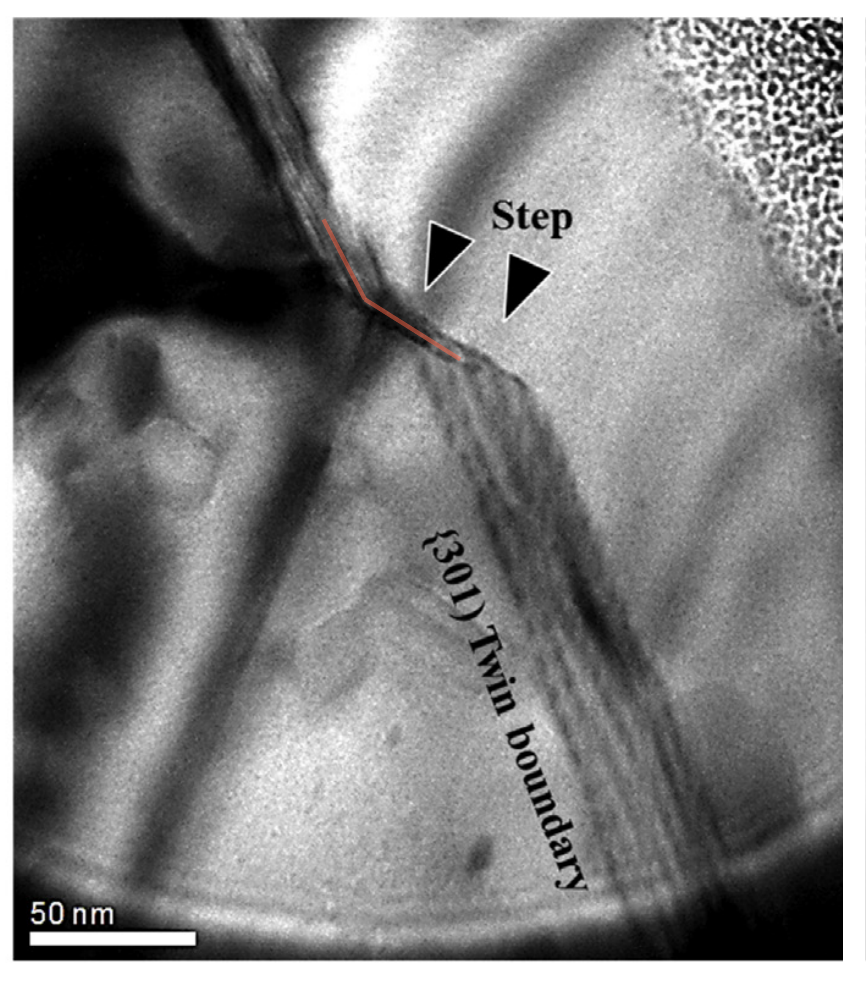}
        \caption{Experimentally observed facet after bicrystal compression \cite{kaira}}
        \label{fig:subexpt}
    \end{subfigure}
    \vspace{.05cm}
    \begin{subfigure}[h]{.75\textwidth}
        \centering
        \includegraphics[width=\textwidth]{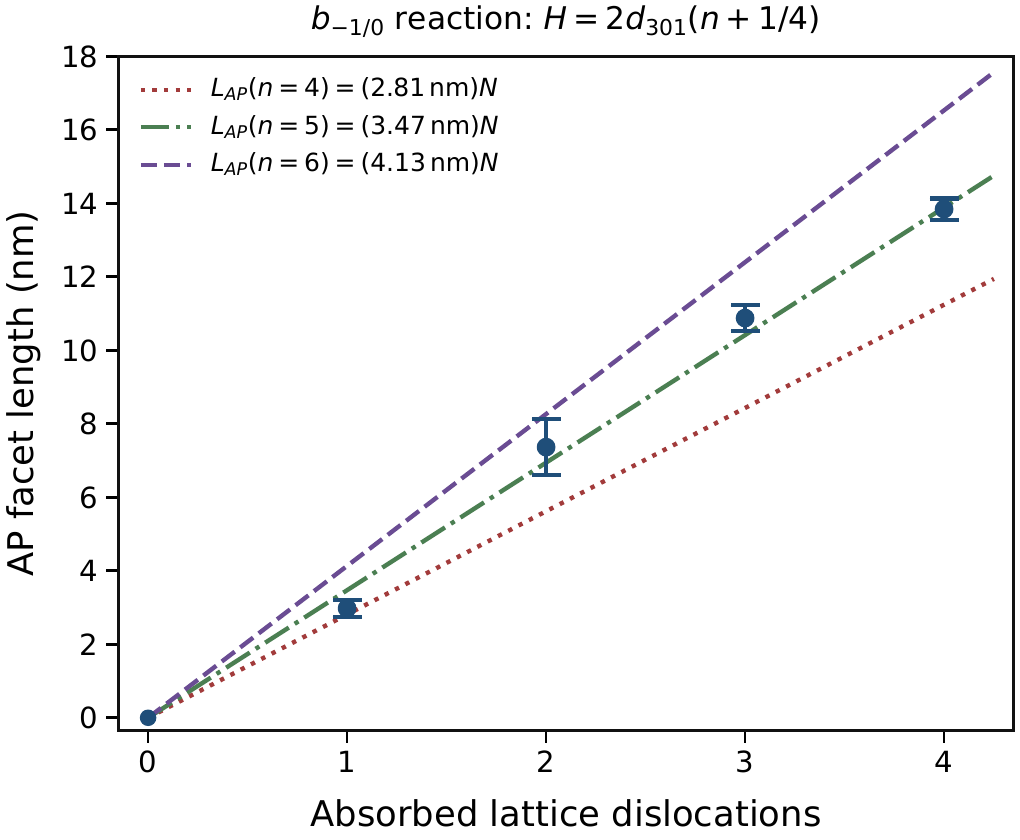}
        \caption{Parameter Survey showing 3 distinct regimes}
        \label{fig:total}
    \end{subfigure}
    \caption{Graph of correlation of number of dislocations colliding with the TB and the resulting facet length. For every number of dislocations there is n=8 measurements}
    \label{fig4:multidislocation}
\end{figure}

Average AP-facet length increases approximately linearly with the number $N$ of absorbed dislocations in our simulations, as shown in Fig.~\ref{fig4:multidislocation}. An automated routine for measuring facet length was developed that (i) isolated the TB core atoms and (ii) measured the length and height of the coherent TB segments. AP facet length was then inferred from the gap required for geometric continuity of the TB segments. The measured reaction is most consistent with $n=5$ geometrically necessary TDs, in good agreement with the $n^*$ estimate from Eq.~2, although the $N=1$ case exhibits smaller-than-average facet lengths with $n$ closer to 4 than 5. The measured AP-facet inclinations, adjusted for simulation-box tilt, vary little with $N$, with a mean value of $30.4^\circ$, between the predicted A- and P-side values ($29.94^\circ$ and $31.72^\circ$). Upon dislocation absorption, the overall GB becomes a faceted asymmetric-tilt boundary. The measured tilt of the coherent twin segment due to accumulating TDs (Supplementary Fig.~S3b) is well fit by a residual step height consistent with $n=5$; the tilt-angle expression is given by Supplementary Eq.~S36. 

The bicrystal compression experiments on \hkl(301) TBs by Kaira et al. \cite{kaira}, reproduced in Fig.~\ref{fig4:multidislocation}d, show striking similarities to our simulation results, including facet formation after dislocation--twin interactions in association with arrays of twinning dislocations and TB migration. The $150^\circ$ marker superimposed in red on the experimentally observed facet in Fig.~\ref{fig4:multidislocation}d strongly suggests AP-facet character, but definitive confirmation would require high-resolution imaging that resolves the atomic structure. Using the $n^*=6$ estimate for experimental lattice parameters, observed facets of 50 nm are predicted to contain approximately a dozen absorbed dislocations and 72 geometrically necessary TDs. Unlike the random GBs studied by Kaira et al. \cite{kaira}, which harden the material relative to the single crystal under compression, the \hkl(301) twin bicrystal yields more easily and fractures sooner than the single crystal, with small strain bursts in the stress-strain curve possibly associated with TB migration events. Our simulations support a picture of TB-mediated plasticity with a small critical stress for twin migration ($\sigma_{YZ} <5$ MPa \cite{Ian}) and twin mobility uninhibited by absorption of \hkl<c> dislocations. Equivalent mobility with increasing facet length is indicated heuristically by the nearly identical position of the TBs after the same simulation time in Fig.~\ref{fig4:multidislocation}a-c.

In addition to the similarities, there are also notable differences between our simulations and the experiments of Kaira et al. \cite{kaira}. At least four slip systems are active in the experiment on one side of the twin, only one of which is \hkl{100}\hkl<001> and which likely contains a mixture of edge/screw character. At the TB, partial dislocation transmission is observed in addition to step formation. Partial transmission involves emission of a leading partial from the TB with a trailing partial remaining pinned or absorbed within the TB. Early theoretical work by Kiho et al. \cite{ishii1963incorporation} predicted the transmutation of a \hkl{100}\hkl<001> dislocation in the twin grain into a $\frac{1}{2}\hkl{101}\hkl<-101>$ partial dislocation in the matrix. We do not observe this reaction for the loading state considered in our simulations, but instead observe \hkl(301) twin nucleation at the lower AP-facet junction. This mechanism is shown in detail in Supplementary Fig.~S1. 

To our knowledge, this work is the first direct atomistic study of TB-dislocation interactions in a tetragonal material, $\beta$-Sn. Strong analogies are found between TB-dislocation interactions in $\beta$-Sn and HCP metals, specifically the well-studied reaction of basal dislocations with the low energy \hkl(10-12) coherent twin to form PB/BP facets in Mg, Ti and Zr \cite{gong2017interface,dang2020formation,hooshmand2017atomistic,yue2023twin}. BP/PB facets in Mg can easily transfer TDs across their step risers by disconnection absorption and emission, allowing the TB to progress through a forest of basal dislocations with little apparent loss of mobility \cite{el2015,dang2020formation}. This behavior has been proposed to contribute to the low growth stress and profuse growth of \hkl{10-12} twins in Mg \cite{el2015,yue2023twin}. Similarly, we observe little change in twin mobility after dislocation absorption and AP facet formation in $\beta$-Sn. This observation helps explain the ubiquity of the \hkl(301) twinning mode in experiments. Faceted TB mobility can be quantified further in future work, including mechanisms of TD transfer across AP facets. 

One notable difference between the \hkl<c>-dislocation--CTB interaction at the \hkl(301) twin in Sn and the basal-dislocation--CTB interaction at the \hkl(10-12) twin in Mg concerns the relative signs of the Burgers vectors associated with unit TDs and facet disclinations. The Sn system has disclination content aligned with the direction of shear coupling, whereas the Mg system does not. This leads to opposite effects on $n^*$. In Mg, the oppositely signed disclination necessitates emission of an additional TD and raises $n^*$ from 4 to 5 \cite{barrett2014faceting,barrett2014roles,el2015}, whereas in Sn, including disclination character in the energy minimization reduces $n^*$ from 6 to 5 (Supplementary Section S2). We note that these reactions are only two examples of the general phenomena of mobile tilt boundary faceting which has been observed across multiple twinning modes in HCP metals \cite{barrett2014faceting}.

MLIP-informed large scale MD simulations of twin-dislocation interactions in tetragonal $\beta$-Sn offer detailed mechanistic insight into the deformation-induced faceting observed in the experiments of Kaira et al. \cite{kaira}. A  dislocation absorption induced faceting transition is demonstrated in a model system, $\hkl<c>$ type edge dislocations interacting with a \hkl(301) CTB. The facet inclination is found to be consistent with low energy AP facets previously characterized in twin growth simulations \cite{Ian}. The participation of $n^*=5$ twinning dislocations in the unit absorption reaction is consistent with calculations from isotropic elasticity theory which account for the disclination character of the AP facet. We hope that these results spur further interest in transmission electron microscopy characterization of deformation induced faceting in $\beta$-Sn; the suggested AP facet structure and TD statistics await direct atomic-scale confirmation. Like \hkl(10-12) twins in HCP Mg, \hkl(301) twins in $\beta$-Sn are hypothesized to be ubiquitous in processed microstructures not only because of their low energy, but also because of their ability to remain mobile (or nucleate as secondary twins) in a forest of dislocations.

\clearpage
\section*{Abbreviations}
\begin{table}[H]
\centering
\small
\caption{Abbreviations used in the main text and Supplementary Information.}
\label{tab:abbreviations}
\begin{tabular}{@{}ll@{}}
\toprule
Abbreviation & Definition \\
\midrule
AP/PA & A-plane--Prismatic and Prismatic--A-plane facet orientations \\
BCT & body-centered tetragonal \\
BP/PB & basal--prismatic and prismatic--basal facet orientations \\
CTB & coherent twin boundary \\
DFT & density functional theory \\
FCC & face-centered cubic \\
GB & grain boundary \\
HCP & hexagonal close-packed \\
LAMMPS & Large-scale Atomic/Molecular Massively Parallel Simulator \\
MD & molecular dynamics \\
MLIP & machine-learned interatomic potential \\
MTP & moment tensor potential \\
NPT & isothermal--isobaric ensemble \\
OVITO & Open Visualization Tool \\
PTM & polyhedral template matching \\
TB & twin boundary \\
TD & twinning disconnection \\
\bottomrule
\end{tabular}
\end{table}

\section*{Acknowledgments}
We acknowledge funding from the Institute for Materials Science at Los Alamos National Laboratory (LANL) through grants 20258285CT-IMS and RR2026-IPD-1.
We also acknowledge funding from the Materials project (ASC-PEM-MAT) of the Physics and Engineering Models subprogram of Advanced Simulation and Computing program at LANL. 

\bibliographystyle{ieeetr}
\bibliography{library}

\clearpage


\section*{Supplementary material}

# Supplementary Information

## Dislocation mediated twin boundary faceting in $\beta$-Sn

John Israel[a,b], Ian Chesser[a], Khanh Dang[c], Vinamra Agrawal[a]

[a] *XCP-5, Los Alamos National Laboratory, Los Alamos, NM USA*
[b] *Department of Aerospace Engineering, Iowa State University, Ames, IA USA*
[c] *MST-8, Los Alamos National Laboratory, Los Alamos, NM USA*

## S1. Secondary twin nucleation

The secondary-twin nucleation mechanism can be viewed as a high-stress extension of the facet-forming absorption reaction discussed in the main text. In Fig. S1a, the dislocation has been absorbed into the $(3\,0\,1)$ twin to form an AP facet ($n = 5$ case discussed in the main text) and migrates with the boundary for several nanometers. Twin nucleation is marked by the loss of linearity of the AP facet as a distorted region develops across the lower twin–AP-facet junction and the lower portion of the facet (Fig. S1b). Upon further twin migration, the nucleated twin fragment is ejected into the lower matrix grain (Fig. S1c), producing a slip trace in the fiducial marker (rightmost arrow in Fig. S1c). This trace is consistent with slip along the $c$ direction in the matrix grain, which could in principle have resulted from direct transmission of the original $\langle c\rangle$ dislocation into that grain. This type of direct reaction is observed at coherent twins in face-centered cubic (FCC) metals and certain hexagonal close-packed (HCP) materials [1, 2, 3] under similar applied stress states. Twin-fragment formation and ejection therefore provide an alternative transmission mechanism with the same net result as direct cross-slip-like transmission. Although polyhedral template matching (PTM) identifies some non-perfect atoms between the twin fragment and the twin boundary (TB) in Fig. S1c, we identify no stacking fault, supporting twin-fragment ejection. Under continued applied stress, the secondary twin grows while the primary twin continues migrating downward (Fig. S1d); the secondary twin is then absorbed into the primary twin, where it continues to grow (Fig. S1e–f).

Dominant twin facets are labeled and include $\{3\,0\,1\}$ and $\{1\,0\,1\}$ facets, as well as AP and Prismatic–A-plane (PA) facets. Angles between $\{3\,0\,1\}$ planes measured across the near-$\{3\,0\,1\}$ and near-$\{1\,0\,1\}$ facets, approximately $61.4^\circ$ and $117.5^\circ$, support a misorientation consistent with $\{3\,0\,1\}$ twins rather than $\{1\,0\,1\}$ twins: they deviate from the ideal coherent TB (CTB) angles reported in the main text by $1.3^\circ$ and $4.1^\circ$, respectively. The $\{3\,0\,1\}$ facets of the secondary twin are closer to coherency than the $\{1\,0\,1\}$ facets, supporting $\{3\,0\,1\}$ twinning.

## S2. Derivation of $n^*$

We proceed to derive the continuous reaction coordinate $n^*$ used in the main text and the geometrical relation used to predict the inclination of the CTB segment. The calculation is a restricted isotropic line-energy model. It includes the elastic self-energy of the translational residual, the independent self-energies of the $n$ isolated $h(2)$ twinning disconnections (TDs), and the signed equivalent Burgers-vector content of the A–P junction disclination dipole. A common elastic prefactor and logarithmic cutoff are assumed for the translational residual and the isolated TDs. AP interfacial energy, coherent

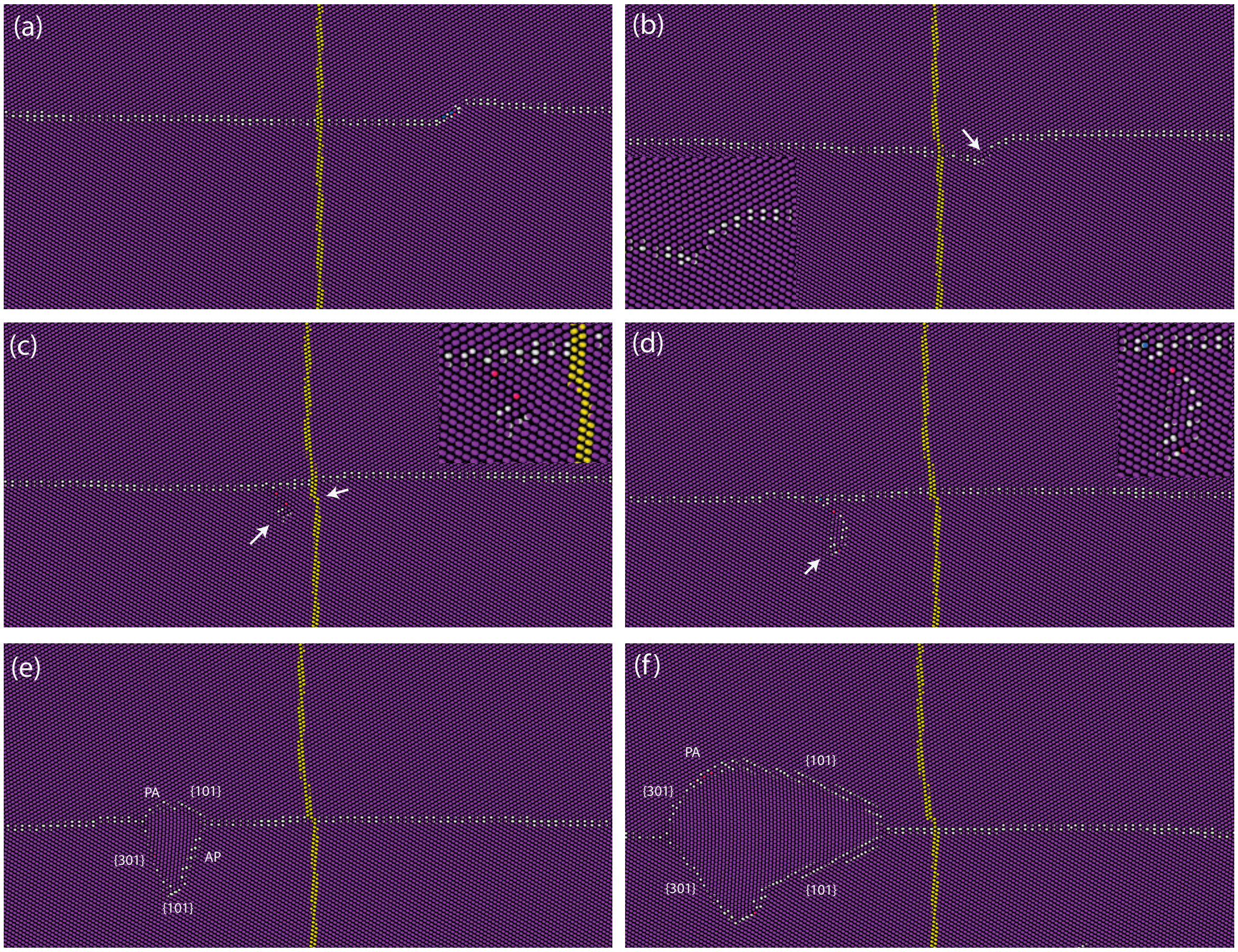


Figure S1: Heterogeneous nucleation of a secondary $\{3\,0\,1\}$ twin at the $(3\,0\,1)$ twin boundary, illustrated at 200 K and 500 MPa resolved shear stress on the original $\{1\,0\,0\}\langle 0\,0\,1\rangle$ slip system in the upper twin grain. A fiducial marker is shown in yellow to track relative displacements during slip and migration. In (a), the dislocation has been absorbed to form an A-plane–Prismatic (AP) facet ($n = 5$ case discussed in the main text). In (b)–(c), a small twin fragment nucleates at the lower AP-facet–TB junction and is ejected into the matrix grain. The insets show enlarged regions indicated by the white arrows. In (d)–(f), the secondary twin grows as it is swept downward by the mobile primary TB. Dominant twin-facet types are labeled.

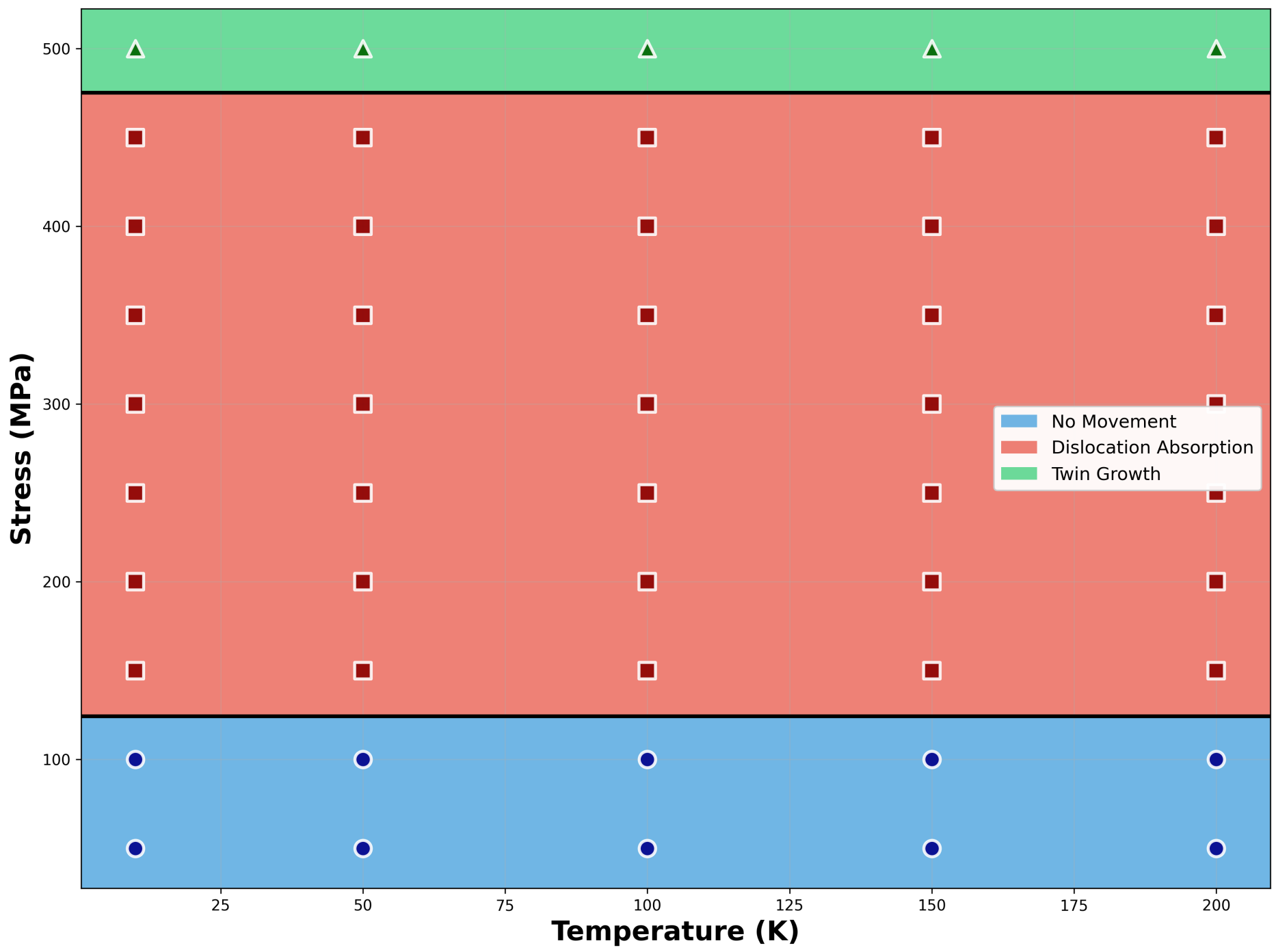


Figure S2: Reaction outcomes for the MD parameter sweep performed in this work reveal no temperature dependence of the apparent critical stress for twin nucleation (green region).

misfit strain, the positive self-energy of the disclination dipole, TD–TD interactions, junction/core energies, applied work, and semicoherent screening are not included. The result should therefore be interpreted as a reproducible reaction estimate rather than a complete reaction-selection theory.

The oriented $b_{p/q}$ notation follows the main text: $p$ and $q$ count the signed numbers of (301) planes terminated in the twin and matrix grains, respectively, and $\boldsymbol{b}_{-p/-q} = -\boldsymbol{b}_{p/q}$. The incident lattice dislocation is $\boldsymbol{b}_{-1/0}$ and the positive elementary TD is $\boldsymbol{b}_{2/2}$. If $n$ positive TDs remain on the boundary, conservation of the $p/q$ indices gives

$$\boldsymbol{b}_{-1/0} = \boldsymbol{b}_{-(2n+1)/-2n} + n\boldsymbol{b}_{2/2}. \tag{S1}$$

For $n = 5$, the residual is $\boldsymbol{b}_{-11/-10}$, consistent with the observed 11 twin/A and 10 matrix/P plane terminations.

Let

$$\gamma = \frac{c}{a}, \qquad \gamma < \frac{1}{\sqrt{3}}, \tag{S2}$$

and let a unit vector tangent to the (301) CTB in the section normal to the dislocation line be

$$\hat{\boldsymbol{t}}_{301} = \frac{(-a, 0, 3c)}{\sqrt{a^2 + 9c^2}}. \tag{S3}$$

In the oriented crystallographic basis used here, the incident lattice translation has representative $\boldsymbol{b}_i = (0, 0, c)$, and the magnitude of the elementary $h(2)$ TD Burgers vector is

$$|\boldsymbol{b}_{2/2}| = a\frac{1 - 3\gamma^2}{\sqrt{1 + 9\gamma^2}}. \tag{S4}$$

Before including the A–P rotational closure, the translational residual is

$$\boldsymbol{b}_R = \boldsymbol{b}_i - n\boldsymbol{b}_{2/2}. \tag{S5}$$

Minimizing only its isotropic self-energy gives the nearest-vector projection

$$n_R = \frac{\boldsymbol{b}_i \cdot \boldsymbol{b}_{2/2}}{|\boldsymbol{b}_{2/2}|^2} = \frac{3\gamma^2}{1-3\gamma^2}. \tag{S6}$$

If the $n$ separated TDs are each assigned the same line-energy prefactor $C > 0$, the restricted energy becomes

$$E_0(n) = C\left(|\boldsymbol{b}_i - n\boldsymbol{b}_{2/2}|^2 + n|\boldsymbol{b}_{2/2}|^2\right). \tag{S7}$$

Differentiation with respect to continuous $n$ gives

$$\frac{1}{2C}\frac{dE_0}{dn} = -\boldsymbol{b}_{2/2}\cdot(\boldsymbol{b}_i - n\boldsymbol{b}_{2/2}) + \frac{1}{2}|\boldsymbol{b}_{2/2}|^2 = 0, \tag{S8}$$

$$n^*_{R+t} = \frac{3\gamma^2}{1-3\gamma^2} - \frac{1}{2}. \tag{S9}$$

The $-1/2$ term is the penalty associated with the independent TD self-energies.

The natural A- and P-plane trace inclinations relative to the CTB are

$$\theta_A(\gamma) = \tan^{-1}\left(\frac{1}{3\gamma}\right), \tag{S10}$$

$$\theta_P(\gamma) = \tan^{-1}\left(\frac{1}{\gamma}\right) - \theta_A(\gamma). \tag{S11}$$

Their mismatch defines the signed A-to-P Frank rotation

$$\phi_{\mathrm{AP}} = \theta_A - \theta_P = 2\tan^{-1}\left(\frac{1}{3\gamma}\right) - \tan^{-1}\left(\frac{1}{\gamma}\right). \tag{S12}$$

For a facet vector $\boldsymbol{L}$, the exact equivalent Burgers-vector content of the two opposite junction disclinations is

$$\boldsymbol{b}_{r,\mathrm{AP}} = \left(\boldsymbol{R}_X[\phi_{\mathrm{AP}}] - \boldsymbol{I}\right)\boldsymbol{L}. \tag{S13}$$

Here $\boldsymbol{R}_X[\phi_{\mathrm{AP}}]$ is the matrix of a right-handed rotation through the Frank angle $\phi_{\mathrm{AP}}$ about the $X$ axis, which is the $[0\,1\,0]$ tilt axis and hence the disclination line direction, and $\boldsymbol{I}$ is the $3\times 3$ identity matrix. The combination $\boldsymbol{R}_X[\phi_{\mathrm{AP}}] - \boldsymbol{I}$ is the standard Frank-vector construction for a wedge disclination dipole: it returns the relative displacement that the rotation $\phi_{\mathrm{AP}}$ generates between the two ends of the facet vector $\boldsymbol{L}$. It vanishes when the A and P traces are parallel ($\phi_{\mathrm{AP}} = 0$), and for small $\phi_{\mathrm{AP}}$ its magnitude grows as $|\boldsymbol{b}_{r,\mathrm{AP}}| \approx |\phi_{\mathrm{AP}}|\,L$, i.e. in proportion to the facet length $L = |\boldsymbol{L}|$. This proportionality is what couples $\boldsymbol{b}_{r,\mathrm{AP}}$ to the facet geometry and forces the self-consistent treatment given below. Taking the A trace as the reference direction gives the signed components

$$\frac{b_{r,\mathrm{AP},Y}}{L} = \cos\theta_A - \cos\theta_P, \qquad \frac{b_{r,\mathrm{AP},Z}}{L} = \sin\theta_P - \sin\theta_A. \tag{S14}$$

The trigonometric factors may be written

$$\cos\theta_A = \frac{3\gamma}{\sqrt{1+9\gamma^2}}, \qquad \sin\theta_A = \frac{1}{\sqrt{1+9\gamma^2}}, \tag{S15}$$

$$\cos\theta_P = \frac{1+3\gamma^2}{\sqrt{1+\gamma^2}\sqrt{1+9\gamma^2}}, \qquad \sin\theta_P = \frac{2\gamma}{\sqrt{1+\gamma^2}\sqrt{1+9\gamma^2}}. \tag{S16}$$

Define

$$q(\gamma) = 3\gamma - \frac{1+3\gamma^2}{\sqrt{1+\gamma^2}}, \tag{S17}$$

so that

$$\cos\theta_A - \cos\theta_P = \frac{q(\gamma)}{\sqrt{1+9\gamma^2}}. \tag{S18}$$

For the Sn lattice-parameter ratios considered here, $q(\gamma) < 0$.

The translational part of the residual AP defect is

$$\boldsymbol{b}_{t,\mathrm{AP}} = \boldsymbol{b}_i - \boldsymbol{b}_{r,\mathrm{AP}} - n\boldsymbol{b}_{2/2}. \tag{S19}$$

Retaining its self-energy and the independent TD self-energies gives

$$E(n,L) = C\left(|\boldsymbol{b}_i - \boldsymbol{b}_{r,\mathrm{AP}} - n\boldsymbol{b}_{2/2}|^2 + n|\boldsymbol{b}_{2/2}|^2\right). \tag{S20}$$

At fixed $L$,

$$\frac{1}{2C}\frac{\partial E}{\partial n} = -\boldsymbol{b}_{2/2}\cdot(\boldsymbol{b}_i - \boldsymbol{b}_{r,\mathrm{AP}} - n\boldsymbol{b}_{2/2}) + \frac{1}{2}|\boldsymbol{b}_{2/2}|^2 = 0, \tag{S21}$$

$$n^*(L) = \frac{\boldsymbol{b}_{2/2}\cdot(\boldsymbol{b}_i - \boldsymbol{b}_{r,\mathrm{AP}})}{|\boldsymbol{b}_{2/2}|^2} - \frac{1}{2}. \tag{S22}$$

Writing the dimensionless facet length as $\lambda = L/a$ and inserting Eqs. (S4) and (S17) reduces the vector projection to

$$\boxed{n^*(\gamma,\lambda) = \frac{3\gamma^2 + \lambda q(\gamma)}{1-3\gamma^2} - \frac{1}{2}.} \tag{S23}$$

The second term in the numerator is the signed A–P closure correction. It lowers $n^*$ for the A-to-P ordering and TD convention used in the main text. Reversing either convention reverses this signed geometrical shift but not the positive self-energy omitted from Eq. (S20).

The (301) plane spacing and elementary double-layer step height are

$$d_{301} = \frac{ac}{\sqrt{a^2+9c^2}} = \frac{a\gamma}{\sqrt{1+9\gamma^2}}, \qquad h(2) = 2d_{301}. \tag{S24}$$

For the residual $\boldsymbol{b}_{-(2n+1)/-2n}$ in Eq. (S1), the magnitude of the net step height is

$$H(n) = 2d_{301}\left(n+\frac{1}{4}\right). \tag{S25}$$

Projection onto the exact A and P traces gives

$$L_A(n) = \frac{H(n)}{\sin\theta_A} = 2c\left(n+\frac{1}{4}\right), \tag{S26}$$

$$L_P(n) = \frac{H(n)}{\sin\theta_P} = \sqrt{a^2+c^2}\left(n+\frac{1}{4}\right). \tag{S27}$$

The main text uses the mean of these exact-side estimates to close the restricted model,

$$L_{AP}(n) = \frac{L_A+L_P}{2} = a\, s_{AP}(\gamma)\left(n+\frac{1}{4}\right), \qquad s_{AP}(\gamma) = \gamma + \frac{1}{2}\sqrt{1+\gamma^2}. \tag{S28}$$

Equation (S23) cannot be evaluated directly, because its right-hand side needs a facet length that is not known until the reaction has been chosen. The dependency runs in a closed loop:

$$n \text{ TDs absorbed } \longrightarrow \text{ step height } H(n), \text{ Eq. (S25) } \longrightarrow \text{ facet length } L_{AP}(n), \text{ Eq. (S28) } \longrightarrow \text{ disclination content } \boldsymbol{b}_{r,\mathrm{AP}} \propto L, \text{ Eq. (S13) } \longrightarrow \text{ optimal } n^*, \text{ Eq. (S23).}$$

Assuming a value of $n$ to fix the facet geometry therefore returns a generally different optimal value $n^*$, and only the value for which the two agree describes a facet that is simultaneously the right length for the number of TDs it contains and the right length to minimize the energy. That value is the fixed point of the loop. Because both the geometric step $L_{AP} \propto (n + \frac{1}{4})$ and the energetic step $n^* \propto \lambda$ are linear in their arguments, the loop closes in closed form and no numerical iteration is required. Define

$$A(\gamma) = \frac{3\gamma^2}{1 - 3\gamma^2} - \frac{1}{2}, \qquad \kappa_{AP}(\gamma) = \frac{s_{AP}(\gamma)q(\gamma)}{1 - 3\gamma^2}. \tag{S29}$$

Substitution of Eq. (S28) into Eq. (S23) gives the linear fixed-point equation

$$n^* = A + \kappa_{AP}\left(n^* + \frac{1}{4}\right), \tag{S30}$$

and hence

$$\boxed{n^{*,\mathrm{sc}}_{AP}(\gamma) = \frac{A(\gamma) + \kappa_{AP}(\gamma)/4}{1 - \kappa_{AP}(\gamma)}.} \tag{S31}$$

The two coefficients have direct interpretations. $A(\gamma)$ is the prediction that would be obtained if the facet carried no disclination content at all, i.e. the translational-plus-TD result of Eq. (S9); it is the entry $n^*_{R+t}$ in Table S1. The coefficient $\kappa_{AP}(\gamma)$ is the gain of the loop above: it is the change in the optimal $n^*$ produced by a unit change in the $n$ assumed when fixing the facet length. For the lattice parameters considered here $q(\gamma) < 0$, so $\kappa_{AP} < 0$ and the feedback is negative ($\kappa_{AP} = -0.26$ for the MTP lattice parameters). Physically, adding a TD lengthens the facet, and the extra disclination content that comes with the extra length supplies part of the closure that would otherwise have to be provided by TDs.

This is the $n^*$ reported in the main text. It is explicitly a function of $a$ and $c$ through

$$\gamma = \frac{c}{a}, \qquad A = \frac{3c^2}{a^2 - 3c^2} - \frac{1}{2}, \tag{S32}$$

$$s_{AP} = \frac{2c + \sqrt{a^2 + c^2}}{2a}, \qquad q = \frac{3c}{a} - \frac{a^2 + 3c^2}{a\sqrt{a^2 + c^2}}, \tag{S33}$$

with $\kappa_{AP} = s_{AP}q/(1 - 3c^2/a^2)$. The physical reaction is selected as the nearest allowed integer.

Table S1: Elasticity predictions before and after including the signed AP disclination closure. Here $n^*_{R+t}$ includes the translational residual and the independent TD self-energies, whereas $n^*_{AP}$ additionally includes the self-consistent rotational content of the AP facet. The moment tensor potential (MTP) and experimental-lattice results use their respective lattice parameters; the latter is a theoretical prediction and not an experimental measurement of $n$.

| Parameters | $a$ (Å) | $c$ (Å) | $n^*_{R+t}$ | $n^*_{AP}$ | selected $n$ |
|---|---|---|---|---|---|
| MTP | 5.97 | 3.22 | 6.358 | 4.983 | 5 |
| Experiment | 5.83 | 3.18 | 7.808 | 6.142 | 6 |

## S3. Measured and predicted CTB facet length and inclination

For $N = 1$ absorbed dislocations, facet length statistics were taken across the entire temperature and applied stress grid consistent with absorption in Fig. S2. For $N = 4$, only two facet length measurements were considered. In between these extremes, intermediate absorption states were leveraged to obtain 6 data points for the $N = 2$ case and 4 data points for the $N = 3$ case.

Let the two endpoints of the AP chord in the deformed $YZ$ section differ by $\Delta Y_{\mathrm{AP}}$ and $\Delta Z_{\mathrm{AP}}$, after periodic unwrapping. The AP segment replaces that projected portion of a periodic boundary whose instantaneous deformed period is $L_y$. The complementary coherent segment must therefore close the height change over the remaining projected length. The measured unsigned CTB closure inclination is

$$\boxed{\phi_{\mathrm{CTB}}^{\mathrm{meas}} = \tan^{-1}\left[\frac{|\Delta Z_{\mathrm{AP}}|}{L_y - |\Delta Y_{\mathrm{AP}}|}\right].} \tag{S34}$$

Equation (S34) is a periodic-box closure angle, not an independent local least-squares fit to the coherent boundary. Using the instantaneous $L_y$ accounts for deformation of the simulation cell.

For comparison with the $n = 5$ P-facet prediction, assume that identical reaction heights accumulate additively after $N$ lattice dislocations are absorbed. Equations (S25) and (S27) give

$$H_N = 2Nd_{301}\left(n + \frac{1}{4}\right) = NL_P \sin\theta_P, \qquad \Delta Y_{P,N} = NL_P\cos\theta_P = \frac{H_N}{\tan\theta_P}. \tag{S35}$$

The predicted P-side closure inclination is consequently

$$\boxed{\phi_{\mathrm{CTB}}^{(P)}(N) = \tan^{-1}\left[\frac{NL_P\sin\theta_P}{L_y(N) - NL_P\cos\theta_P}\right] = \tan^{-1}\left[\frac{H_N}{L_y(N) - H_N/\tan\theta_P}\right].} \tag{S36}$$

The dashed curve in Fig. S3b uses $n = 5$, the MTP lattice parameters, the exact P inclination, and the mean deformed value of $L_y(N)$ for each absorption state.

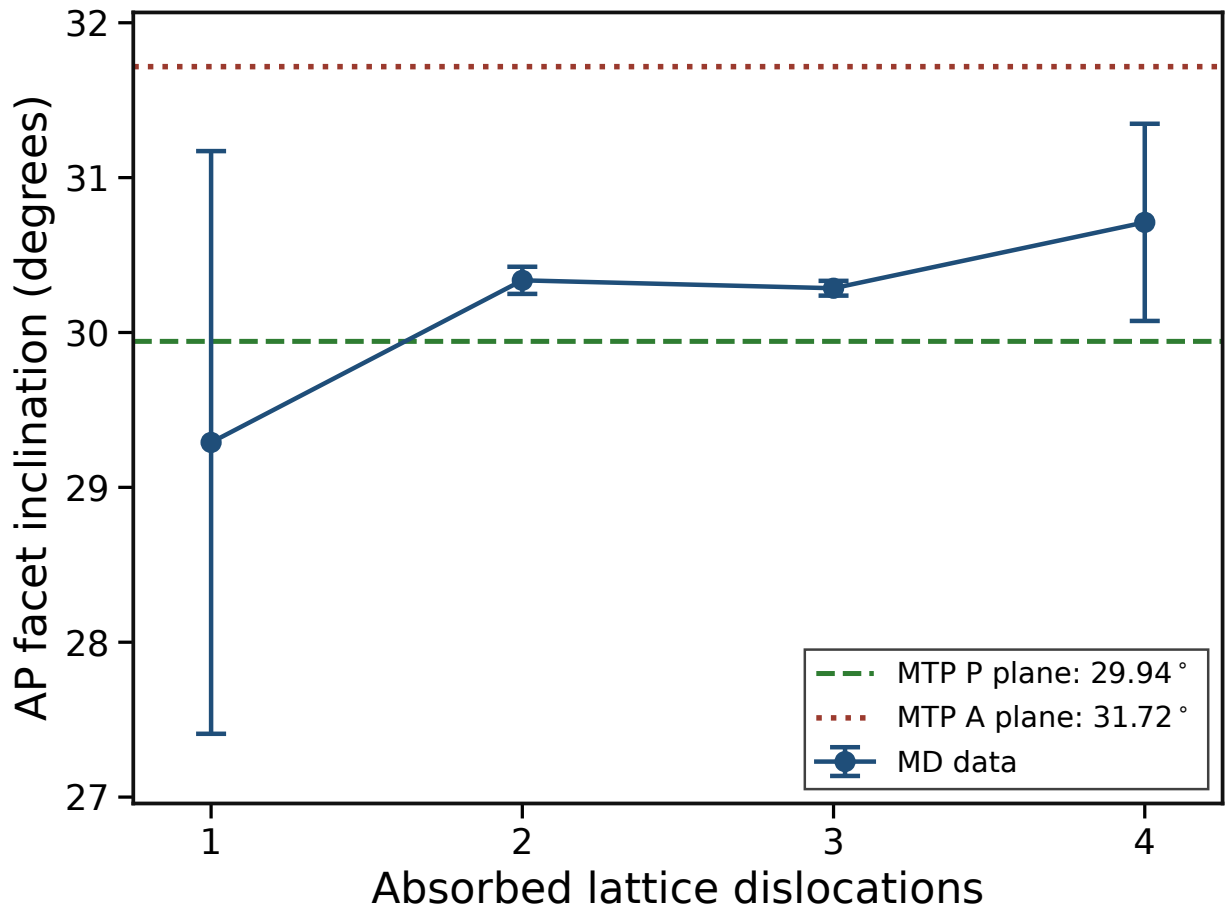


(a) AP-facet inclination.

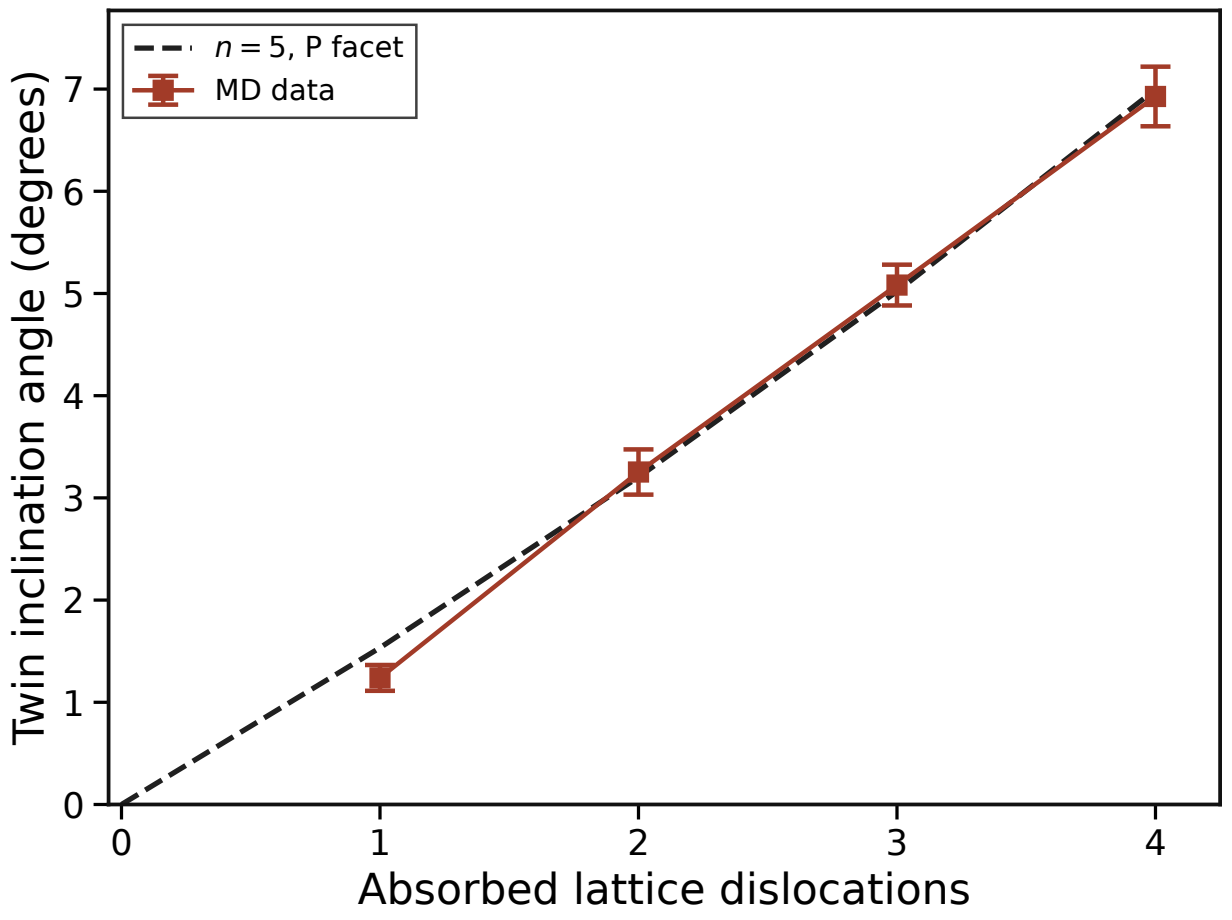


(b) Coherent twin-segment inclination.

Figure S3: Inclination statistics after one through four absorbed lattice dislocations. (a) The AP inclination remains close to the ideal MTP P-plane value and varies weakly with absorption count. (b) The global inclination of the complementary coherent segment increases with accumulated residual height and agrees closely with Eq. (S36) for $n = 5$ and the exact P-facet projection. Error bars show the campaign-level standard deviation used in the main-text analysis.


\end{document}